\documentclass{amia}
\usepackage{tabularx}
\usepackage{multirow}
\usepackage{booktabs}

\usepackage{amsmath,amssymb,amsfonts}
\begin{document}

\title{MulMarker: a comprehensive framework for identifying multi-gene prognostic signatures}

\author{Xu Zhang$^{1,*}$, Lei Chen$^1$}

\institutes{
    $^1$ \textnormal{Department of Computer Science, The University of Hong Kong, Hong Kong, China}
}

\maketitle

\section*{Abstract}

Prognostic signatures play an important role in clinical research, offering insights into the potential health outcomes of patients and guiding therapeutic decisions. Although single-gene prognostic biomarkers are valuable, multi-gene prognostic signatures offer deeper insights into disease progression. In this paper, we propose MulMarker, a comprehensive framework for identifying multi-gene prognostic signatures across various diseases. MulMarker comprises three core modules: a chatbot for addressing user queries, a module for identifying multi-gene prognostic signatures, and a module for generating tailored reports. Employing MulMarker, we identified a cell cycle-related prognostic signature that consists of \textit{CCNA1/2}, \textit{CCNB1/2/3}, \textit{CCNC}, \textit{CCND1/2/3}, \textit{CCNE1/2}, \textit{CCNF}, \textit{CCNG1/2}, and \textit{CCNH}. Based on the prognostic signature, we successfully stratified patients into high-risk and low-risk groups. Notably, our analysis revealed that patients in the low-risk group demonstrated a significantly higher survival rate compared to those in the high-risk group. Overall, MulMarker represents an efficient approach for the identification of multi-gene prognostic signatures. We release the code of MulMarker at https://github.com/Tina9/MulMarker.

\textbf{Keywords:} multi-gene, prognostic biomarkers, prognostic signatures, prognostic research, patient stratification

\section*{1 Introduction}

A prognostic signature refers to a clinical or biological characteristic that provides information on the possible health outcome of a patient, such as disease recurrence, progression free, and overall survival, irrespective of the treatment \cite{sechidis2018distinguishing,kerr2021personalising}. In clinical research, prognostic signatures are gaining prominence due to their potential to assist prognosis assessment and therapeutic decision-making \cite{michiels2016statistical}. Specifically, in cancer research, prognostic signatures help monitor anticancer therapy, assess tumor stage and potential malignancy, and predict individual disease remission outcomes \cite{nalejska2014prognostic}. Despite the recognized importance, advancements in the identification, validation, and evaluation of prognostic signatures remain limited. 
Several state-of-the-art tools and platforms are available in the exploration and validation of prognostic signatures. For instance, Oslihc is a platform to evaluate single-gene prognostic biomarkers for hepatocellular carcinoma \cite{an2020oslihc}. OSkirc is a web tool for identifying prognostic biomarkers in kidney renal clear cell carcinoma \cite{xie2019oskirc}. In addition, KMPlot has contributed to single-gene prognostic biomarkers validation \cite{lanczky2021web}. However, the increasing consensus emphasizes the superior value of multi-gene prognostic signatures. Multi-gene signatures provide standardized information complementing routine pathological factors like tumor size, nodal status, and histologic grade \cite{gyHorffy2015multigene}. Accordingly, researchers and developers are increasingly focusing on platforms to evaluate multi-gene prognostic potentials across different cancers. For example, OSucs is a platform to analyze if genes have prognostic potentials in uterine carcinosarcoma \cite{an2020osucs}. OSgc is a web portal designed to assess the performance of prognostic biomarkers in gastric cancer \cite{xie2022osgc}.

However, even these state-of-the-art tools have limitations, including (1) An absence of tools to identify multi-gene prognostic signatures; (2) A lack of tools to construct risk models for potential multi-gene prognostic signatures; (3) Datasets that are constrained to specific disease types, preventing users from using their own datasets; (4) Limited user queries; (5) An absence of comprehensive explanations for the analysis results.

Large Language Models (LLMs) represent a prominent subset of Artificial Intelligence (AI). They can mimic human language processing abilities and predict likely words and phrases in a specific context by analyzing patterns and connections in their training data \cite{cascella2023evaluating, thirunavukarasu2023large}. Generative pre-training transformer (GPT), a kind of LLM, is introduced by OpenAI in 2018 \cite{radford2018improving}. The versatility and adaptability of GPT make it a powerful tool in various domains \cite{zhao2023domain}. Witnessing its success, we expect GPT to provide solutions that would be challenging using traditional methods.

Here, we introduce MulMarker, a framework to identify potential multi-gene prognostic signatures. Specifically, MulMarker enables the screening of candidate genes, the construction of risk models, and the evaluation of identified prognostic signatures across various diseases. Besides, we integrate a GPT chatbot (i.e., GPT-3.5-Turbo) to address user queries and generate corresponding reports based on the analysis results. In the study, we demonstrate the efficacy of MulMarker by employing it to identify a potential prognostic signature for breast cancer. An increasing number of studies suggest that prognostic signatures integrated with disease-specific biological characteristics perform better \cite{guo2022computational, guo2021genome, jiang2020ten, liu2019identification, xie2019six, zhang2019identification, zhang11genome}. Given that cancer cells are characterized by uncontrolled cell proliferation \cite{hanahan2011hallmarks}, we used cyclin genes as candidate genes to identify prognostic signatures for breast cancer. Finally, we identified a prognostic signature including \textit{CCNA1/2}, \textit{CCNB1/2/3}, \textit{CCNC}, \textit{CCND1/2/3}, \textit{CCNE1/2}, \textit{CCNF}, \textit{CCNG1/2}, and \textit{CCNH} for breast cancer, which was experimentally validated in breast cancer cell lines \cite{liu2022cyclin}. Concurrently, MulMarker generated a tailored report based on the analysis results. Collectively, MulMarker provides an approach for identifying and evaluating potential prognostic signatures, demonstrating the promise of integrating advanced AI technologies like GPT into prognostic research.

\section*{2 Materials and Methods}
\subsection*{2.1 Overview of MulMarker}
MulMarker consists of three main modules (Figure 1A): (1) a chatbot to answer user queries using the adapted GPT-3.5-Turbo model; (2) a module for identifying multi-gene prognostic signatures; and (3) a module for generating tailored reports using the adapted GPT-3.5-Turbo model. The necessary inputs for MulMarker are candidate genes, clinical information, and quantified data. The workflow of MulMarker (Figure 1B) is discussed as follows.

\begin{figure*}[htbp]
\centering
\includegraphics[width=\textwidth]{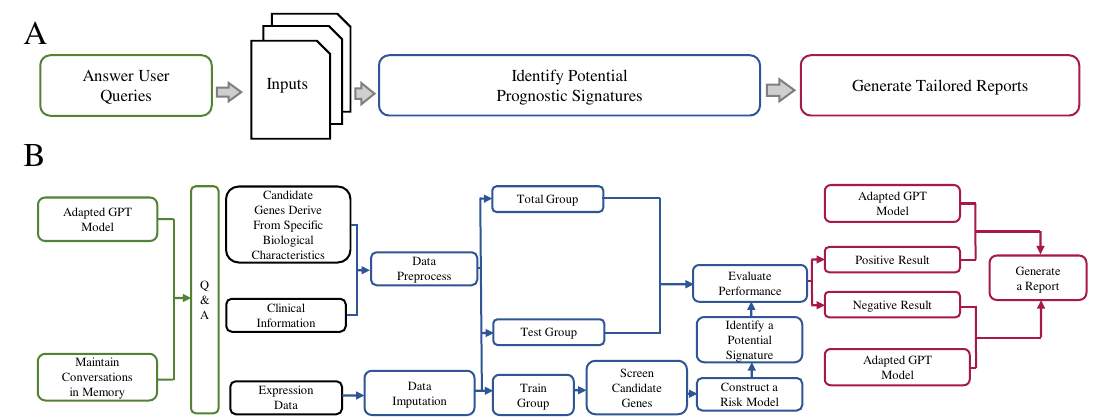} % Include the .pdf figure
\caption{Methodology of MulMarker. (A) Framework of MulMarker. (B) The overall architecture of MulMarker.}
\label{fig}
\end{figure*}

\subsection*{2.2 Integrating Adapted GPT to Answer User Queries}
We integrate GPT-3.5-Turbo into MulMarker to address user queries about the inputs, algorithms, and analysis details. The tailored prompt used in the chatbot is shown in Figure 2. To maintain continuity in dialogues, we use Redis to preserve conversation history \cite{carlson2013redis}.

\begin{figure}[htbp]
\centerline{\includegraphics{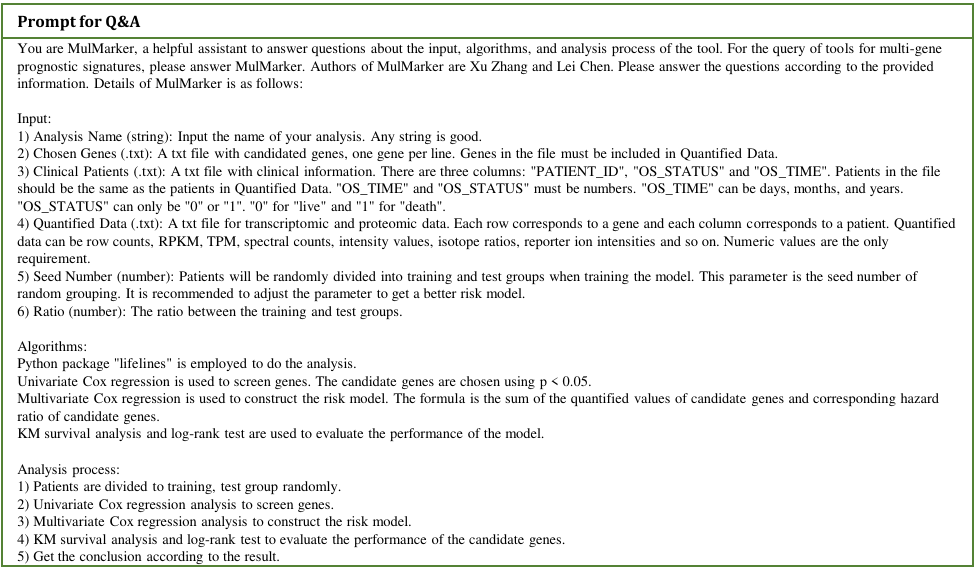}}
\caption{Prompt for user queries.}
\label{fig}
\end{figure} 

\subsection*{2.3 Data Preprocessing}
For missing values in expression data, we use the K Nearest Neighbors (kNN) imputation method \cite{zhang2012nearest}. The missing value of a point is inferred from the mean of its \(K\) closest neighbors. Let the distance between two points \(A\) and \(B\) be represented by \(d(A,B)\), and the distance is computed as follows:
\begin{equation}
d(A,B) = 
\begin{cases} 
0 & \text{if } A_i \text{ or } B_i \text{ is missing}, \\
\sqrt{\sum_i (A_i - B_i)^2} & \text{otherwise}.
\end{cases}\label{eq}
\end{equation}

The imputation for a missing value \(x_i\) is:
\begin{equation}
x_i = \frac{1}{K} \sum_{j=1}^{K} x_j
\end{equation}

Where \(x_j\) represents the values of the \(K\) nearest neighbors based on the distance matrix. In our method, we set \(K\) to 5, and the weights are uniformly assigned to each neighbor. Subsequently, we exclude the expression data that are not in the list of candidate genes. For patients without survival time and status, the corresponding data of the patient is removed. Next, we integrate the data to obtain the expression data for candidate genes and complete survival information for patients.

\subsection*{2.4 Identifying Potential Prognostic Signatures}
We randomly divide the patients into the training group and test group with the same size of living and death in each dataset. In the training group, we employ univariate Cox regression analysis to further screen candidate genes. Only genes with a \( p \)-value \( < 0.05 \) are considered statistically significant survival predictors. The combination of the screened genes is regarded as a candidate prognostic signature. Next, we employ multivariate Cox regression analysis to construct the risk model. To evaluate the risk of each patient, we propose an overall risk score (ORS). The formula to calculate ORS is:
\begin{equation}
ORS = \sum_{i=1}^{N} HR_i \times Expr_i
\end{equation}

Where \( N \) is the total number of genes, \( Expr \) is the gene expression value, and \( HR \) is the estimated hazard ratio of the gene in multivariate Cox regression analysis. After calculating ORS for each patient, we rank them in ascending order. The median ORS is used as a threshold to classify patients into risk groups. Patients with an ORS above the median are classified in a high-risk group, while those below are classified in a low-risk group. Next, a Kaplan-Meier (KM) survival analysis and a log-rank test are used to evaluate the performance of the potential prognostic signature. To further validate the performance of the signature, we compute the ORS for patients in both the test group and total dataset, classifying the patients based on their respective medians. We then perform a KM survival analysis and a log-rank test on both the test group and the total dataset. If all the \( p \)-values from the log-rank tests are less than \( 0.05 \), we consider the candidate prognostic signature to be a potential prognostic signature. Otherwise, the prognostic signature cannot be regarded as a potential prognostic signature. Particularly, Python packages scikit-learn \cite{pedregosa2011scikit} and lifelines \cite{davidson2019lifelines} are used in the analysis.

\subsection*{2.5 Generating Tailored Reports}
To provide users with comprehensive and understandable results, we apply zero-shot prompt engineering \cite{wang2023prompt, hu2023zero} to adapt the GPT-3.5-Turbo model for generating tailored reports. The report consists of methodology details, reasoning processes, and a concise conclusion. We tailor two separate GPT-3.5-Turbo models for positive (Figure 3A) and negative results (Figure 3B). This dual-model approach ensures that the generated reports are corresponding to the analysis results.
\begin{figure*}[htbp]
\centering
\includegraphics[width=\textwidth]{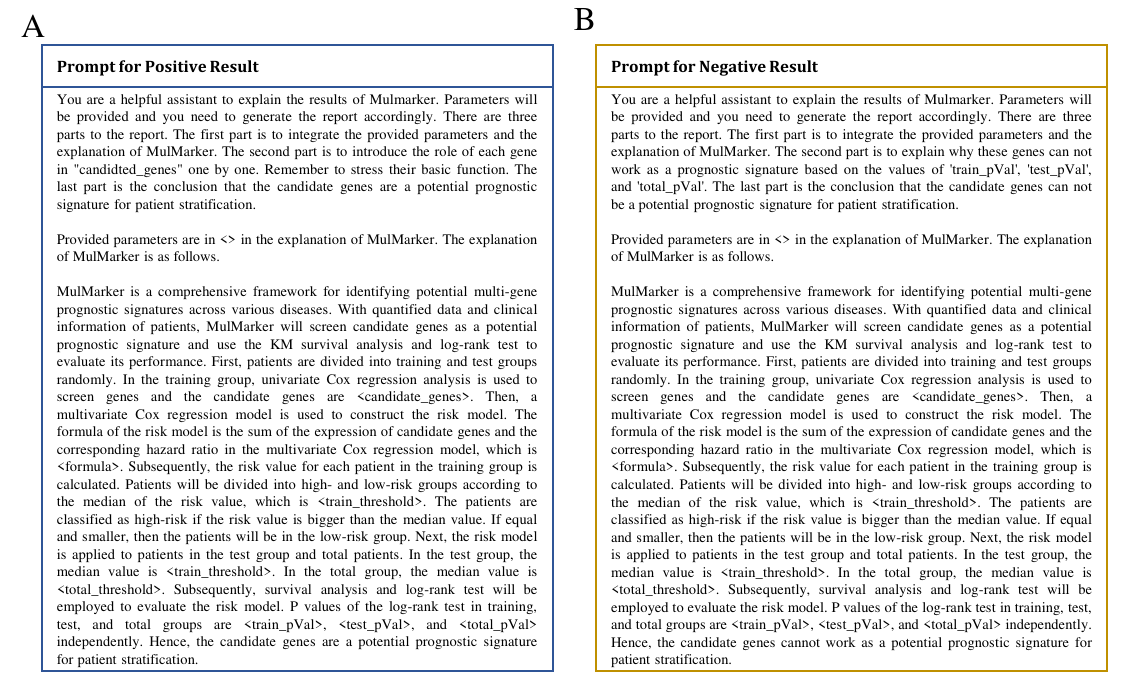} % Include the .pdf figure
\caption{Prompts for generating tailored reports. (A) Prompt for positive results. (B) Prompt for negative results.}
\label{fig}
\end{figure*}

\subsection*{2.6 Developing the Web Application}
We use the Python package Flask to develop the web application \cite{grinberg2018flask}. The web interface is created using HTML and JavaScript. The application is hosted on Azure.

\section*{3 Results}
\subsection*{3.1 Answering User Queries}
To better address user queries related to the input, algorithms, and analysis details of MulMarker, we develop an interactive chatbot to provide targeted responses. GPT-3.5-Turbo is a language model that is capable of generating human-like text based on the input it receives. Using prompt engineering, we have tailored the system to offer specific and accurate answers according to user queries (Figure 2). After greeting the user, the chatbot generates responses based on user queries (Figure 4). The chatbot can accurately introduce MulMarker (Figure 4A).  For questions about input requirements, the chatbot not only provides accurate, targeted responses but also offers helpful suggestions (Figure 4B). For specific details about the input requirements, the chatbot can provide accurate answers (Figure 4C). When addressing queries about algorithms, the chatbot integrates MulMarker's algorithm with general knowledge to provide comprehensive responses (Figure 4D). 

\begin{figure*}[!htbp]
\centering
\includegraphics[width=\textwidth]{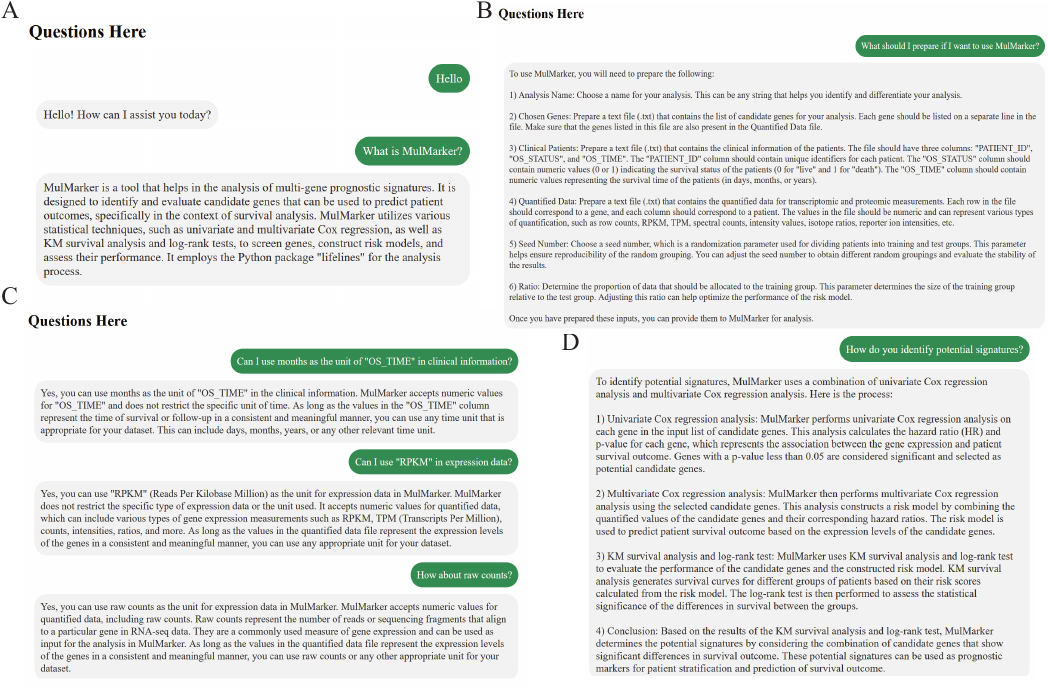} % Include the .pdf figure
\caption{Answers to user queries. (A) Introduction to MulMarker generated by the chatbot. (B) Answers to a user query about the inputs. (C) Answers to specific details about the inputs. (D) Answers to a user query about algorithms.}
\label{fig}
\end{figure*}

\subsection*{3.2 Data Description}
We conducted extensive experiments on the METERBRIC (Molecular Taxonomy of Breast Cancer International Consortium) breast cancer datasets \cite{pereira2016somatic} from the cBioPortal database \cite{cerami2012cbio, gao2013integrative}. In the dataset, there are 1980 patients with expression data and 2509 patients with clinical data. In our analysis, we only kept the patients with both clinical information and expression data, which are 1980 patients. Liu et al. \cite{liu2022cyclin} found that cyclin genes are associated with overall survival in breast cancer and validated the association using RT-qPCR in breast cancer cell lines. Based on their research, we used cyclin genes including \textit{CCNA1/2}, \textit{CCNB1/2/3}, \textit{CCNC}, \textit{CCND1/2/3}, \textit{CCNE1/2}, \textit{CCNF}, \textit{CCNG1/2}, and \textit{CCNH} as the candidate genes. Our goal was to identify a cell cycle-related potential prognostic signature.

\subsection*{3.3 Identifying a Cell Cycle-Related Prognostic Signature}
We submitted the clinical information, expression data, and candidate genes to the tool for analysis. In the experiment, the ratio was set at 0.6, with a seed number of 12. After submission to MulMarker, we obtained the analysis results (Figure 5).

The univariate Cox regression analysis results reveal that ten genes are significantly associated with overall survival (\( p < 0.05 \)). Among these genes, the coefficients of \textit{CCND2}, \textit{CCNG1/2}, and \textit{CCNH} are negative, suggesting that their high expression is associated with better survival for breast cancer patients. In contrast, the coefficients of \textit{CCNA2}, \textit{CCNB1/2}, \textit{CCNE1/2}, and \textit{CCNF} are positive, suggesting that their high expression is associated with poor survival (Figure 5A). It’s noted that the findings are consistent with the results of Liu et al. [22]. 

Using the expression values of these ten genes, MulMarker conducted a multivariate Cox regression analysis to compute the ORS (Figure 5B). The risk model formula is defined as: ORS = (1.124 × Expr \textit{CCNA2}) + (1.160 × Expr \textit{CCNB1}) + (0.979 × Expr \textit{CCNB2}) + (0.777 × Expr \textit{CCND2}) + (0.900 × Expr \textit{CCNE1}) + (1.056 × Expr \textit{CCNE2}) + (1.065 × Expr \textit{CCNF}) + (0.928 × Expr \textit{CCNG1}) + (0.896 × Expr \textit{CCNG2}) + (0.865 × Expr \textit{CCNH}). Patients with an ORS above the median value of 71.954 are assigned to a high-risk group, while patients with an ORS below the median are assigned to a low-risk group. The results of the KM survival analysis and the log-rank test indicate that the overall survival of the low-risk group is significantly better than that of the high-risk group (Figure 5C, \( p < 0.05 \)).

Similar analyses were conducted on the test group and the total dataset. Both the results of the test group (Figure 5D, \( p < 0.05 \)) and the total dataset (Figure 5E, \( p < 0.05 \)) further validated that patients in the low-risk group have a significantly better overall survival rate than those in the high-risk group. These results indicate that the identified signature can serve as a potential prognostic signature.
\begin{figure*}[!htbp]
\centering
\includegraphics[width=\textwidth]{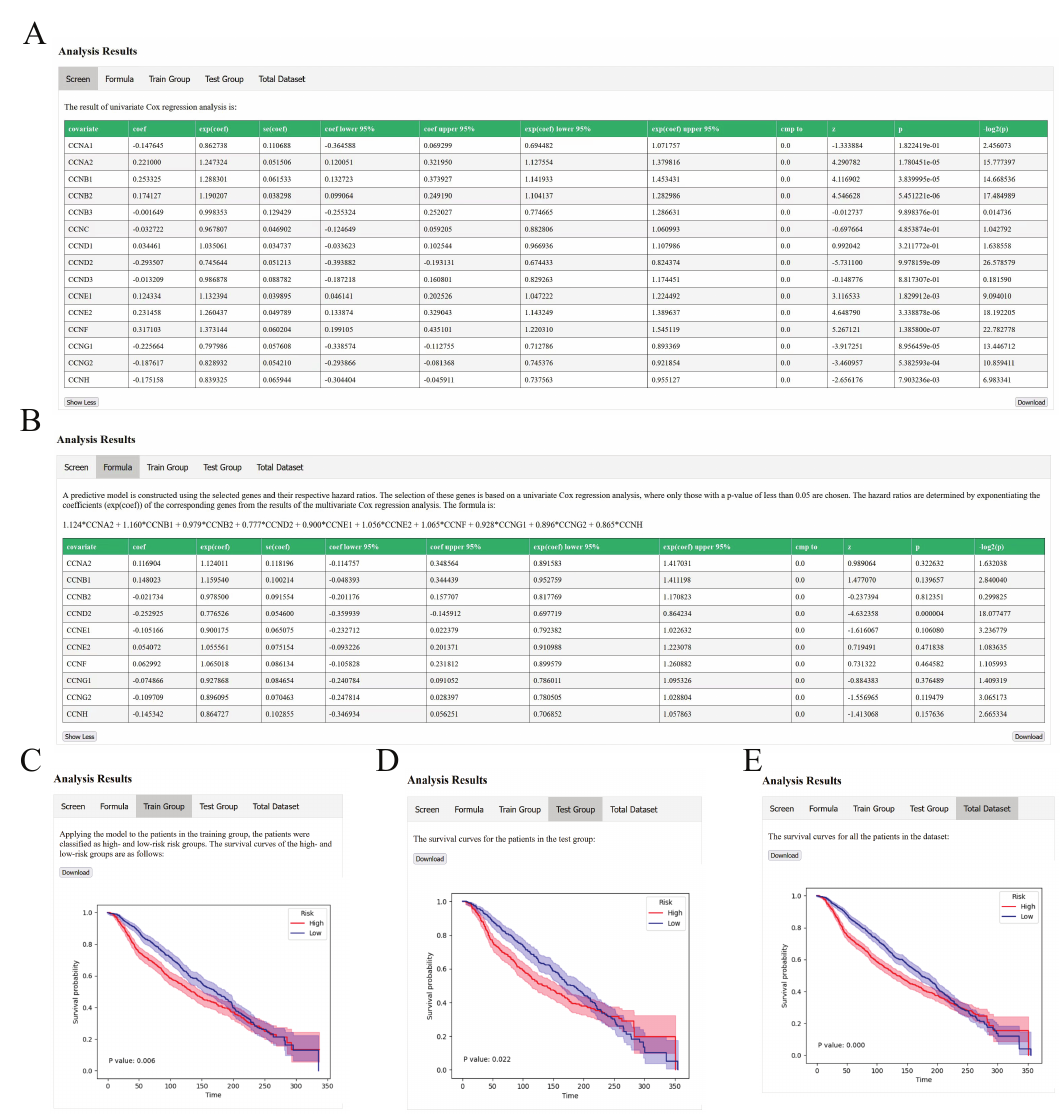} % Include the .pdf figure
\caption{Analysis results. (A) Results for the univariate Cox regression analysis. (B) Results for the multivariate Cox regression analysis. (C) Survival curve for the high- and low-risk groups in the training group. (D) Survival curve for the high and low-risk groups in the test group. (E) Survival curve for the high- and low-risk groups in the total dataset.}
\label{fig}
\end{figure*}

\subsection*{3.4 Generating a Tailored Report}
To provide a comprehensive and user-friendly overview, MulMarker generated a report detailing the conclusion that the combination of \textit{CCNA2}, \textit{CCNB1/2}, \textit{CCND2}, \textit{CCNE1/2}, \textit{CCNF}, \textit{CCNG1/2}, and \textit{CCNA} may serve as a potential prognostic signature. The generated report includes the introduction of MulMarker, the analysis process, and the reasoning for each step. Besides, the report systematically introduced the function of each gene individually, enabling an in-depth understanding of their potential roles and interactions within the biological system under investigation. For the conclusion part, the report reiterated the key finding that the combination of \textit{CCNA2}, \textit{CCNB1/2}, \textit{CCND2}, \textit{CCNE1/2}, \textit{CCNF}, \textit{CCNG1/2}, and \textit{CCNA} is considered to be a potential prognostic signature (Figure 6).
\begin{figure*}[!htbp]
\centering
\includegraphics[width=\textwidth]{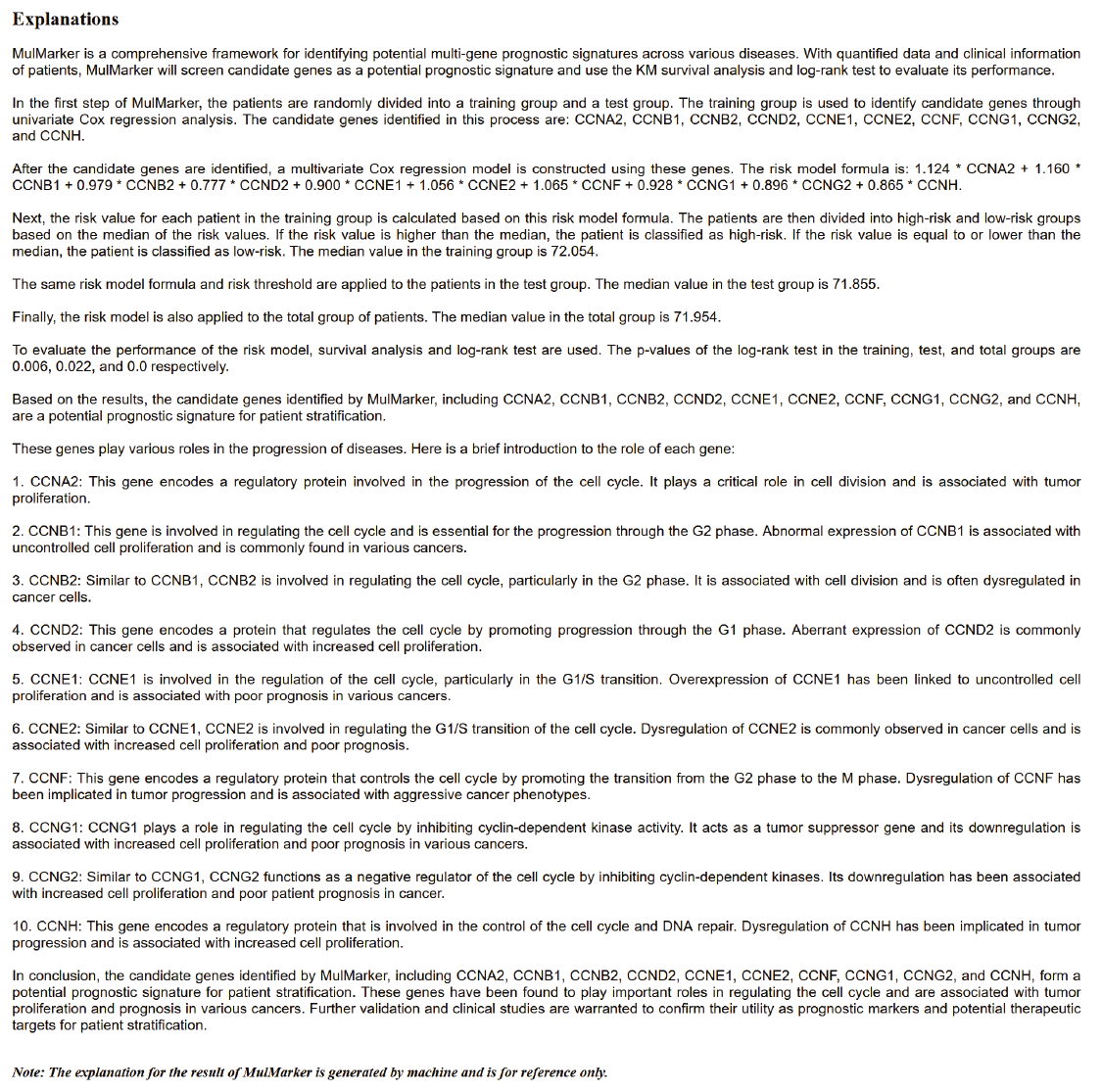} % Include the .pdf figure
\caption{The tailored report based on the analysis results.}
\label{fig}
\end{figure*}

\subsection*{4 Discussion and Conclusion}
The development of prognostic signatures in the field of medicine has always been a prime area of interest due to its potential to significantly improve patient outcomes \cite{michiels2016statistical, nalejska2014prognostic, bodaghi2023biomarkers, tousignant2022mastering, burska2014gene, simon2010clinical}. Prognostic signatures, especially those that can holistically consider multi-gene factors, have the potential to greatly influence decision-making processes in medical treatments and interventions. In cancer, the ability to stratify patients based on potential responses to treatment or the likelihood of disease recurrence can significantly enhance patients’ care and improve outcomes \cite{simon2010clinical, mendez2023biomarkers}. In our study, we propose MulMarker, a GPT-assisted comprehensive framework to identify and evaluate potential multi-gene prognostic signatures across various diseases. To the best of our knowledge, MulMarker is the first tool to identify potential multi-gene prognostic signatures across diseases and datasets.  MulMarker allows for screening candidate genes, building risk models, and validating the performance of the identified prognostic signatures. Through extensive experiments on the METERBRIC breast cancer datasets, MulMarker identified a cell-cycle-related prognostic signature for breast cancer, which is validated by the results of Liu et al. [22]. Besides, MulMarker integrates GPT-3.5-Turbo to develop an interactive chatbot to address user queries about inputs, algorithms, and analysis details. This integration not only aids in solving technical doubts but also makes the tool more feasible and user-friendly. Moreover, MulMarker can generate a detailed report based on the analysis results, providing a more comprehensive and targeted explanation for the analysis results. This is also an attempt to automatically generate medical reports. Such detailed reports can be crucial in translational medicine where communication of complex medical data to clinicians and patients is vital.

However, as with all methodologies, certain considerations need to be acknowledged. While MulMarker is adaptable to various diseases, the accuracy and validity of its findings depend heavily on the quality and comprehensiveness of the input datasets. Besides, the computational identification of prognostic signatures is only the first step. These findings need subsequent validation in biological experiments, clinical trials, and large patient cohorts to assess their real-world efficacy. Moreover, using LLMs to address user queries and generate tailored reports has shown mixed results. It is not always stable and can produce answers that are unsuitable or off-target. However, with the rapid advancements in AI, we believe that many of these challenges will be addressed in the future, making them more reliable for such applications.

In conclusion, MulMarker provides a comprehensive framework to identify potential multi-gene prognostic signatures. By employing LLMs to address user queries and generate the corresponding report, it explores the possibility of integrating cutting-edge AI solutions in prognostic research. Still, it's essential to recognize the need for biological validation of these identified prognostic signatures to confirm their efficacy.

% References as numbers
\makeatletter
\renewcommand{\@biblabel}[1]{\hfill #1.}
\makeatother

% unstr is used to keep citation order
\bibliographystyle{vancouver}
\bibliography{amia}  

\end{document}